\documentclass[12pt]{article}

\usepackage[height=8.85in,width=6.45in]{geometry}

\usepackage[utf8]{inputenc}
\usepackage[T1]{fontenc}
\usepackage{amsmath}
\usepackage{amssymb}
\usepackage{mathtools}
\numberwithin{equation}{section}
\allowdisplaybreaks

\usepackage{slashed}
\usepackage{braket}
\usepackage[usenames,dvipsnames,svgnames,table]{xcolor}
\usepackage[colorlinks,citecolor=DarkGreen,linkcolor=FireBrick,urlcolor=FireBrick,linktocpage]{hyperref}
\usepackage{cite}
\usepackage{graphicx}
\usepackage{tikz}
\usepackage{tikz-cd}
\usepackage{times}
\usepackage[scaled]{couriers}

\usepackage{bm}
\usepackage{subfig}

\usepackage{xcolor}
\usepackage{mdframed}

\renewenvironment{figure}[1][]{
  \begin{originalfigure}[#1]
    \begin{mdframed}[linecolor=black!0,backgroundcolor=black!1]
}{
    \end{mdframed}
  \end{originalfigure}
}

\definecolor{identifiercolor}{rgb}{.4,.6,.56}
\definecolor{stringcolor}{gray}{0.5}
\definecolor{inactivecolor}{rgb}{0.15,0.15,0.5}
\usepackage{listings}
\def\vev#1{\langle#1\rangle}
\def\code#1{\texttt{#1}}
\def\inv#1{\mathop{\mathrm{inv}\vev{#1}}}
\def\Irr{\mathop{\mathrm{Irr}}}
\def\id{\mathop{\mathrm{id}}}

\newcommand{\op}[1]{\mathcal{#1}}
\newcommand{\abs}[1]{\left\lvert#1\right\rvert}

\begin{document}

\begin{titlepage}

\begin{flushright}
IPMU-19-0022
\end{flushright}

\vskip 3cm

\begin{center}

{\LARGE \code{autoboot}}

\bigskip
\bigskip

{\Large \bfseries a generator of bootstrap equations
with global symmetry}

\bigskip
\bigskip

Mocho Go and Yuji Tachikawa

\bigskip

\begin{tabular}{ll}
 & Kavli Institute for the Physics and Mathematics of the Universe (WPI), \\
& University of Tokyo, Kashiwa, Chiba 277-8583, Japan
\end{tabular}

\vskip 1cm

\end{center}

\noindent
We introduce \code{autoboot},
a \code{Mathematica} program which automatically generates mixed correlator bootstrap equations of an arbitrary number of scalar external operators, given the global symmetry group and the representations of the operators.
The output is a \code{Python} program which uses Ohtsuki's \code{cboot} which in turn uses Simmons-Duffin's \code{sdpb}. The code is available at \url{https://github.com/selpoG/autoboot/}.

In an appendix we also discuss a simple technique to significantly reduce the time to run \code{sdpb}, which we call hot-starting.

\end{titlepage}

\setcounter{tocdepth}{2}
\tableofcontents

\section{Introduction and summary}

A four-point function $\vev{\phi_1\phi_2\phi_3\phi_4}$ in a conformal field theory (CFT) can be constructed from three-point functions, but in more than one way, depending on how to group the four operators for the operator product expansions (OPEs): $(\phi_1\phi_2)(\phi_3\phi_4)$ or $(\phi_1\phi_3)(\phi_2\phi_4)$ or $(\phi_1\phi_4)(\phi_2\phi_3)$.
The bootstrap equation expresses the equality of the four-point function computed in these different decompositions, and is one of the fundamental consistency conditions of a conformal field theory.

The bootstrap equation has been known for almost a third of a century, see e.g.~\cite{Dobrev:1975ru,Dobrev:1977qv};
for other early papers, we refer the reader to the footnote 4 of \cite{Poland:2018epd}.
It is particularly powerful in 2d where the conformal group is infinite dimensional,
and was already successfully applied for the study of 2d CFT in 1984 in the paper \cite{Belavin:1984vu}.
Its application to CFTs in dimension higher than two had to wait until 2008,
where the seminal paper \cite{Rattazzi:2008pe} showed that a clever rewriting into a form where linear programming was applicable allowed us to extract detailed numerical information from the bootstrap equation.
The technique was rapidly developed by many groups and has been applied to many systems.
A sample of references includes \cite{Rychkov:2009ij,Caracciolo:2009bx,Poland:2010wg,Rattazzi:2010gj,Rattazzi:2010yc,Vichi:2011ux,Poland:2011ey,Rychkov:2011et,ElShowk:2012ht,Liendo:2012hy,ElShowk:2012hu,Beem:2013qxa,Kos:2013tga,Alday:2013opa,Gaiotto:2013nva,Berkooz:2014yda,El-Showk:2014dwa,Nakayama:2014lva,Nakayama:2014yia,Alday:2014qfa,Chester:2014fya,Kos:2014bka,Caracciolo:2014cxa,Nakayama:2014sba,Bae:2014hia,Beem:2014zpa,Chester:2014gqa,Simmons-Duffin:2015qma,Bobev:2015jxa,Kos:2015mba,Chester:2015qca,Beem:2015aoa,Iliesiu:2015qra,Poland:2015mta,Lemos:2015awa,Lin:2015wcg,Chester:2015lej,Chester:2016wrc,Nakayama:2016cim,Behan:2016dtz,Nakayama:2016jhq,Iha:2016ppj,Kos:2016ysd,Nakayama:2016knq,Echeverri:2016ztu,Li:2016wdp,Lin:2016gcl,Bae:2016yna,Bae:2016jpi,Lemos:2016xke,Beem:2016wfs,Li:2017ddj,Cornagliotto:2017dup,Nakayama:2017vdd,Dymarsky:2017xzb,Chang:2017xmr,Cuomo:2017wme,Keller:2017iql,Cho:2017fzo,Li:2017kck,Dymarsky:2017yzx,Bae:2017kcl,Dyer:2017rul,Chang:2017cdx,Cornagliotto:2017snu,Agmon:2017xes,Rong:2017cow,Baggio:2017mas,Stergiou:2018gjj,Hasegawa:2018yqg,Liendo:2018ukf,Rong:2018okz,Atanasov:2018kqw,Behan:2018hfx,Kousvos:2018rhl,Cappelli:2018vir,Gowdigere:2018lxz,Li:2018lyb,Karateev:2019pvw}.\footnote{%
Here we only listed numerical works directly extending and/or using the approach of \cite{Rattazzi:2008pe}.
The authors are afraid that some of the references are inadvertently missing from the list;
they apologize in advance for the omission, and will be happy to include any additional references when notified.
}

Some of the highlights in these developments for the purposes of this paper are:
the consideration of constraints from the global symmetry in \cite{Rattazzi:2010yc},
the introduction of the semi-definite programming in \cite{Poland:2011ey},
and the extension of the analysis to the mixed correlators in \cite{Kos:2014bka}.
These techniques can now constrain the scaling dimensions of operators of 3d Ising and $\mathrm{O}(N)$ models within precise islands.

By now, we have a plethora of good introductory review articles on this approach, see e.g.~\cite{El,Qualls:2015qjb,Rychkov:2016iqz,Simmons-Duffin:2016gjk,Antunes:2017vbx,Poland:2018epd},
which make the entry to this fascinating and rapidly growing subject easier.
There are also various computing tools developed to perform the numerical bootstrap more easily and more efficiently.
For example, we now have a dedicated semi-definite programming solver \code{sdpb} \cite{Simmons-Duffin:2015qma},
and two \code{Python} interfaces to \code{sdpb}, namely
\code{PyCFTBoot} \cite{Behan:2016dtz}
and \code{cboot} \cite{Nakayama:2016jhq}.
There is also a \code{Julia} implementation \code{JuliBootS} \cite{Paulos:2014vya},
and also a \code{Mathematica} package to generate 4d bootstrap equations of arbitrary spin \cite{Cuomo:2017wme}.

Given that the numerical bootstrap of precision islands of the 3d Ising and $O(N)$ models \cite{Kos:2014bka} was done in 2014,
it would not have been strange if there had been many papers studying CFTs with other global symmetries.
But this has not been the case, with a couple of exceptions e.g.~\cite{Rong:2017cow,Baggio:2017mas,Stergiou:2018gjj,Kousvos:2018rhl}.
We believe that this dearth of works concerning CFTs with global symmetry is due to the inherent complexity in writing down the mixed-correlator bootstrap equations, with the constraints coming from the symmetry.
To solve this problem we need to automate it: human beings should never do things which can be done by machines.

The aim of this paper is to present \code{autoboot}, a proof-of-concept implementation of an automatic generator of mixed-correlator bootstrap equations of scalar operators with global symmetry.
Let us illustrate the use with an example.
Suppose we would like to perform the numerical bootstrap of a CFT invariant under $D_8$, the dihedral group with eight elements.
Let us assume the existence of two scalar operators, one in the singlet and the second in the doublet of $D_8$.
The mixed-correlator bootstrap equations can be generated by the following \code{Mathematica} code, after loading our package:
\begin{lstlisting}
	g=getGroup[8,3];
	setGroup[g];
	setOps[{op[e,rep[1]], op[v,rep[5]]}];
	eq=bootAll[];
	sdp=makeSDP[eq];
	py=toCboot[sdp];
	WriteString["D8.py",py];
\end{lstlisting}
Let us go through the example line by line:
\begin{enumerate}
\item \lstinline{g=getGroup[8,3]} sets the group $D_8$ to $g$.
This line illustrates the ability of \code{autoboot} to obtain the group theory data from the \code{SmallGrp} library \cite{SmallGrp} of the computer algebra system \code{GAP} \cite{GAP},
which contains the necessary data of finite groups of order less than $2000$ and many others.
The pair $(8,3)$ is a way to specify a finite group in \code{SmallGrp}.
It simply says that $D_8$ is the third group in their list of groups of order 8.
\item \lstinline{setGroup[g]} tells \code{autoboot} that the symmetry group is $D_8$.
\item \lstinline{setOps[...]} adds operators to \code{autoboot}.
\lstinline{rep[n]} means the $n$-th representation of the group in the \code{SmallGrp} library;
we set the operator \lstinline{e} to be a singlet and the operator \lstinline{v} to be a doublet.
\item \lstinline{eq=bootAll[]} creates the bootstrap equations in a symbolic form and sets them into \lstinline{eq}.
\item \lstinline{sdp=makeSDP[eq]} converts the bootstrap equations into the form of a semi-definite programming problem.
\item \lstinline{py=toCboot[sdp]} further rewrites it into an actual \code{Python} program, which uses \code{cboot} \cite{Nakayama:2016jhq} which internally uses \code{Sage} \cite{Sage}.
\item The last line simply writes the \code{Python} code into an external file.
\end{enumerate}

All what remains is to make a small edit of the resulting file, to set up the dimensions and gaps of the operators.
The \code{Python} code then generates the XML input file for \code{sdpb}.

\bigskip

The rest of the paper is organized as follows.
In Sec.~\ref{sec:theory}, we first explain our notations for the group theory constants
and then describe how the bootstrap equations can be obtained, given the set of external scalar primary operators $\phi_i$ in the representation $r_i$ of the symmetry group $G$.
In Sec.~\ref{sec:implementation}, we discuss how our \code{autoboot} implements the procedure given in Sec.~\ref{sec:theory}.
In Sec.~\ref{sec:examples}, we describe two examples of using \code{autoboot}.
The first is to perform the mixed-correlator bootstrap of the 3d Ising model.
The second is to study the $O(2)$ model with three types of external scalar operators.
Without \code{autoboot}, it is a formidable task to write down the set of bootstrap equations,
but with \code{autoboot}, it is immediate.

We also have an Appendix~\ref{sec:hot} where we discuss a simple technique, which we call \emph{hot-starting}, to reduce the running time of the semi-definite program solver significantly, by reusing parts of the computation for a given set of scaling dimensions of external operators to the computation of another nearby set of scaling dimensions. Our experience shows that it often gives an increase in the speed by about a factor of 10 to 20.

The authors hope that our \code{autoboot} will be of use to the bootstrap community.
The code is freely available at \url{https://github.com/selpoG/autoboot/}.

\section{Theory}
\label{sec:theory}
\subsection{Group theory notations}
Let us first set up our notation for the group theory data we need.
Let $G$ be the symmetry group we are interested,
and $\Irr(G)$ be the set containing one explicit representation
for each isomorphism class of unitary irreducible representations of $G$.

In particular, $r\in \Irr(G)$ is a vector space
together with explicit unitary matrices $U(g)^a_b$ where $a,b=1,\ldots, \dim(r)$
representing the $G$ action: \begin{equation}
v^a \mapsto U(g)^a_b v^b.
\end{equation}
The complex conjugate representation $r^*$ has the $G$ action given by \begin{equation}
v^{a*} \mapsto \overline{U(g)}{}^{a*}_{b*}v^{b*}.
\end{equation}

For $r\in \Irr(G)$, we denote by $\bar r$ the irreducible representation in $\Irr(G)$ isomorphic to $r^*$, i.e.~ $ r^*\simeq \bar r \in \Irr(G)$.
When $r$ is strictly real or complex, we can and do require that $\bar r=r^*$.
When $r$ is pseudoreal, we have $r=\bar r\neq r^*$.
This subtle distinction between $r^*$ and $\bar r$ are unfortunately necessary, since we will carry out the computations using explicit representation matrices.

We denote the $G$-invariant subspace of the tensor product of $r_1,\ldots,r_n$ by $\inv{r_1,\ldots,r_n}$.
We then define $\inv{r_1,\ldots,r_n | s_1,\ldots, s_m}$ to be $\inv{ r^*_1,\ldots, r^*_n, s_1,\ldots, s_m}$.
In particular, we are interested in $\inv{t|r,s}$, whose orthonormal basis times $\sqrt{\dim(t)}$ we denote by \begin{equation}
\Set{\frac{c}{t}|\frac{a,b}{r,s}}_n
\end{equation} where $n=1,\ldots,\dim\inv{t|r,s}$ and
$a,b,c$ are the indices for the irreducible representation $r$, $s$ and $t$.
These are the (generalized) Clebsch-Gordan coefficients of the group $G$.
The scaling factor $\sqrt{\dim(t)}$ is introduced so that we have the relation
\begin{equation}
	\sum_{ab}\overline{\Set{\frac{c}{t}|\frac{a,b}{r,s}}_n}\Set{\frac{c'}{t'}|\frac{a,b}{r,s}}_{n'}
	=\delta_{tt'}\delta_{nn'}\delta_{cc'}.
\end{equation}
When $r \not\simeq s$, we can choose the bases so that \begin{equation}
	\Set{\frac{c}{t}|\frac{a,b}{r,s}}_n=\Set{\frac{c}{t}|\frac{b,a}{s,r}}_n.
\end{equation}
When $r = s$, we can choose signs $\sigma_n(t|r,r)=\pm 1$ so that \begin{equation}
	\Set{\frac{c}{t}|\frac{a,a'}{r,r}}_n
	=\sigma_n(t|r,r)\Set{\frac{c}{t}|\frac{a',a}{r,r}}_n.
	\label{eq:com12}
\end{equation}

We note that $\dim\inv{r|r,\id}=1$ and \begin{equation}
	\Set{\frac{a'}{r}|\frac{a,1}{r,\id}}_1=\delta_{a'a}
\end{equation} for general $r$.
We define the invariant tensor $\Set{a,b}_r$ by \begin{equation}
	\Set{a,b}_r := \sqrt{\dim(r)} \Set{\frac{1}{\id}|\frac{a,b}{r,\bar r}}_1.
\end{equation}
When $r\neq \bar r$ or $r$ is strictly real, we can choose $\Set{a,b}_r$ to be $\delta_{ab}$.
When $r$ is pseudoreal, $\Set{a,b}_r$ is antisymmetric and can be taken to be the direct sum of $\begin{pmatrix}
0 & -1 \\
1 & 0
\end{pmatrix}.$

After these preparations, we can finally write down the orthonormal basis of $\inv{r,s,t}$ and $\inv{r_1,r_2,r_3,r_4}$:
\begin{align}
	\Braket{\frac{a,b,c}{r,s,t}}_n
	&=\frac{1}{\sqrt{\dim(t)}}\sum_{\bar{c}}\Set{\frac{\bar{c}}{\bar{t}}|\frac{a,b}{r,s}}_n
	\Set{\bar{c},c}_{\bar{t}},\\
	\Braket{\frac{a_1,a_2,a_3,a_4}{r_1,r_2,r_3,r_4}}_{s;nm}
	&=\frac{1}{\sqrt{\dim(s)}}\sum_{b\bar{b}}
	\Set{\frac{b}{s}|\frac{a_1,a_2}{r_1,r_2}}_n
	\Set{\frac{\bar{b}}{\bar{s}}|\frac{a_3,a_4}{r_3,r_4}}_m\Set{b,\bar{b}}_{s}.
	\label{aaaa}
\end{align}

We note that
\begin{equation}
	\Braket{\frac{a,b,c}{r,s,t}}_n=\sigma_n(r,s,t)\Braket{\frac{b,a,c}{s,r,t}}_n
\end{equation}
where $\sigma_n(r,s,t):=\sigma_n(\bar t| r,s)$.
Other permutations are more complicated.
We define $\tau$ for the cyclic permutation:
\begin{equation}
	\sum_m\tau_{nm}(r,s,t)\Braket{\frac{a,b,c}{r,s,t}}_n
	=\Braket{\frac{b,c,a}{s,t,r}}_m
\end{equation}
which can be computed via
\begin{align}
	\tau_{nm}(r,s,t)
	&=\sum_{abc}\Braket{\frac{a,b,c}{r,s,t}}_n^*
	\Braket{\frac{b,c,a}{s,t,r}}_m.
\end{align}
We define $\omega$ for the complex conjugation:
\begin{equation}
	\sum_m\omega_{nm}(r,s,t)\Braket{\frac{a,b,c}{r,s,t}}_n
	=\sum_{\bar{a}\bar{b}\bar{c}}
	\Set{\bar{a},a}_{\bar{r}}\Set{\bar{b},b}_{\bar{s}}\Set{\bar{c},c}_{\bar{t}}
	\Braket{\frac{\bar{a},\bar{b},\bar{c}}{\bar{r},\bar{s},\bar{t}}}_m^*
\end{equation}
which can be computed via
\begin{equation}
	\omega_{nm}(r,s,t)
	=\sum_{abc\bar{a}\bar{b}\bar{c}}\Braket{\frac{a,b,c}{r,s,t}}_n^*
	\Set{\bar{a},a}_{\bar{r}}\Set{\bar{b},b}_{\bar{s}}\Set{\bar{c},c}_{\bar{t}}
	\Braket{\frac{\bar{a},\bar{b},\bar{c}}{\bar{r},\bar{s},\bar{t}}}_m^*.
\end{equation}
For the four-point functions, we have obvious relations
\begin{align}
	\Braket{\frac{a_1,a_2,a_3,a_4}{r_1,r_2,r_3,r_4}}_{s;nm}
	&=\sigma_n(s|r_1,r_2)\Braket{\frac{a_2,a_1,a_3,a_4}{r_2,r_1,r_3,r_4}}_{s;nm}\\
	&=\sigma_m(\bar s|r_3,r_4)\Braket{\frac{a_1,a_2,a_4,a_3}{r_1,r_2,r_4,r_3}}_{s;nm}.
\end{align}
\def\six{
\begin{tikzpicture}[baseline=(0)]
\node (n) at  (-30pt,+30pt) {$n$};
\node (m) at  (+30pt,-30pt) {$m$};
\node (0) at (0,0) {};
\node (k) at  (+30pt,+30pt) {$k$};
\node (l) at  (-30pt,-30pt) {$l$};
\draw[thick,-] (n)-- node[below, pos=0.2]{$\vphantom{t}s$}  (m);
\draw[thick,-] (k)-- node[below, pos=0.2]{$t$}  (l);
\draw[thick,-] (n)-- node[pos=.5,below] {$r_1$} (k);
\draw[thick,-] (k)-- node[pos=.5,right] {$r_4$} (m);
\draw[thick,-] (m)-- node[pos=.5,above] {$r_3$} (l);
\draw[thick,-] (l)-- node[pos=.5,left] {$r_2$} (n);
\end{tikzpicture}
}

The final nontrivial relation is
\begin{equation}
	\sum_{s;nm}\six
	\Braket{\frac{a_1,a_2,a_3,a_4}{r_1,r_2,r_3,r_4}}_{s;nm}
	=\Braket{\frac{a_1,a_4,a_3,a_2}{r_1,r_4,r_3,r_2}}_{t;kl}
\end{equation}
which can be solved as
\begin{align}
	\six&=\sum_{a_1a_2a_3a_4}\Braket{\frac{a_1,a_2,a_3,a_4}{r_1,r_2,r_3,r_4}}_{s;nm}^*
	\Braket{\frac{a_1,a_4,a_3,a_2}{r_1,r_4,r_3,r_2}}_{t;kl}
\end{align}
which can be further written as the sum of products of four generalized CG coefficients $\displaystyle\Set{\frac{c}{t}|\frac{a,b}{r,s}}_n$ using \eqref{aaaa}.
This explains our use of the tetrahedron for the coefficients in this relation.
For $G=SU(2)$, this tetrahedral object is known as the 6j symbol.

\subsection{Operator product expansions}
\label{sec:OPE}
For two scalar primary fields $\phi_{1,2}$, we denote its operator product expansion (OPE) by \begin{equation}
	\phi_{1,r[a]}(x)\phi_{2,s[b]}(y)=\sum_{\op{O}:t}
	\sum_{n=1}^{\dim\inv{t|r,s}}\lambda_{\phi_1\phi_2\op{O}}^n
	\Set{\frac{c}{t}|\frac{a,b}{r,s}}_n
	C_{\phi_1\phi_2\op{O},k}\left(x-y,\partial_y\right)\op{O}_{t[c]}^k(y)
\end{equation}
Here, the subscript $r[a]$ means that the operator belongs to the representation $r$ with the index $a=1,\ldots, \dim(r)$,
$\op{O}:t$ means that the intermediate primary operator $\op{O}$ is in the representation $t$,
the superscript $k$ is an index of a spin-$\ell$ representation of the rotation group $SO(d)$,
and $C_{\phi_1\phi_2\op{O},k}(x-y,\partial_y)$ captures the contribution of the descendants.

For an operator $\op{O}$ transforming in $r\in \Irr(G)$, we denote its complex conjugate by $\bar{\op{O}}$ in the representation $\bar r$.
We normalize the operators so that they have the two-point function
\begin{equation}
	\Braket{\op{O}^i_{r[a]}(x)\op{O}'^{j}_{s[b]}(0)}=\delta_{\op{O}\bar{\op{O}}'}
	\Set{a,b}_r\frac{\sigma(\op{O})I^{ij}(x)}{\abs{x}^{2\Delta}}
\end{equation}
where $I^{ij}(x)$ is a certain invariant tensor.
We also introduced signs $\sigma(\op{O})=\pm1$ to compensate the antisymmetry of $\Set{a,b}_r$ when $r$ is pseudoreal.
Namely, if $\op{O}$ is in a complex or in a strictly real representation, $\sigma(\op{O})=+1$,
and for a pair $\op{O}$, $\bar{\op O}$ of operators in a pseudoreal representation,
we choose the sign so that $\sigma(\op{O})\sigma(\bar{\op{O}})=-1$.

We can now proceed to three-point functions. Using the OPE, we find
\begin{multline}
	\Braket{\phi_{1,r[a]}(x_1)\phi_{2,s[b]}(x_2)\op{O}^i_{t[c]}(x_3)}\\
	=\sum_n\alpha_{\phi_1\phi_2\op{O}}^n\Braket{\frac{a,b,c}{r,s,t}}_n
	\frac{\sigma(\bar{\op{O}})Z^i(x)}{\abs{x_{12}}^{\Delta_1+\Delta_2-\Delta}
	\abs{x_{23}}^{\Delta_2+\Delta-\Delta_1}\abs{x_{31}}^{\Delta+\Delta_1-\Delta_2}}
\end{multline}
where $Z^i(x)$ is a certain invariant tensor and we introduced
\begin{equation}
	\alpha_{\phi_1\phi_2\op{O}}^n:=
	\lambda_{\phi_1\phi_2\bar{\op{O}}}^n\sqrt{\dim{t}}.
\end{equation}

The OPE coefficients $\alpha_{\phi_1\phi_2\op{O}}$ have various symmetries.
Firstly, when the spin of $\op{O}$ is $l$, we have \begin{equation}
	\alpha_{\phi_1\phi_2\op{O}}^n
	=\sigma_n(r,s,t)(-1)^l\alpha_{\phi_2\phi_1\op{O}}^n.
	\label{eq:ope12}
\end{equation}
Secondly, when $\op{O}^i_{t[c]}$ is a primary scalar $\phi_{3,r_3[c]}$,
we have
\begin{equation}
	\alpha_{\phi_1\phi_2\phi_3}^n
	=\sigma(\bar{\phi}_1)\sigma(\bar{\phi}_3)
	\sum_m\tau_{nm}(r_1,r_2,r_3)\alpha_{\phi_2\phi_3\phi_1}^m.
	\label{eq:opetau}
\end{equation}
Thirdly, for the complex conjugation, we have
\begin{equation}
	\alpha_{\phi_1\phi_2\op{O}}^n=\sigma(\bar{\phi}_1)\sigma(\bar{\phi}_2)\sigma(\op{O})
	\sum_m\omega_{nm}(r,s,t)(\alpha_{\bar{\phi}_1\bar{\phi}_2\bar{\op{O}}}^{m})^*.
	\label{eq:opeomega}
\end{equation}

The OPE coefficients satisfy the relations \eqref{eq:ope12}, \eqref{eq:opetau}, \eqref{eq:opeomega}.
In particular, when $\op{O}$ is an unknown intermediate operator, we have various relations among
\begin{equation}
	\alpha_{12\op{O}}^n,\alpha_{21\op{O}}^n,
	\alpha_{\bar{1}\bar{2}\bar{\op{O}}}^n,
	\alpha_{\bar{2}\bar{1}\bar{\op{O}}}^n,
\end{equation}
for $n=1,\ldots,\dim\inv{r,s,t}$.
When $\op{O}$ is one of the known external scalar operator $\phi_3$,
there are various relations among
\begin{equation}
	\alpha_{123}^n,\alpha_{231}^n,
	\alpha_{312}^n,\alpha_{213}^n,
	\alpha_{321}^n,\alpha_{132}^n,
	\alpha_{\bar{1}\bar{2}\bar{3}}^n,\alpha_{\bar{2}\bar{3}\bar{1}}^n,
	\alpha_{\bar{3}\bar{1}\bar{2}}^n,\alpha_{\bar{2}\bar{1}\bar{3}}^n,
	\alpha_{\bar{3}\bar{2}\bar{1}}^n,\alpha_{\bar{1}\bar{3}\bar{2}}^n.
\end{equation}

These relations are all $\mathbb{R}$-linear.
Therefore, the solutions to these relations can be parameterized by
mutually independent real numbers we call $\beta_{12\op{O}}^m$,
so that all the OPE coefficients listed above
are linear combinations thereof.

\subsection{Bootstrap equations}
\label{sec:boot}
We can finally study the four-point function.
From the definitions we have given so far, we have
\begin{multline}
	\Braket{\phi_{1,r_1[a_1]}(x_1)\phi_{2,r_2[a_2]}(x_2)
	\phi_{3,r_3[a_3]}(x_3)\phi_{4,r_4[a_4]}(x_4)}\\
	=
	\frac{1}{\abs{x_{12}}^{\Delta_1+\Delta_2}\abs{x_{34}}^{\Delta_3+\Delta_4}}
	\left(\frac{\abs{x_{24}}}{\abs{x_{14}}}\right)^{\Delta_{12}}
	\left(\frac{\abs{x_{14}}}{\abs{x_{13}}}\right)^{\Delta_{34}}\\
	\times\sum_{\op{O}:s}\sum_{nm}\sigma(\op{O})\alpha_{\phi_1\phi_2\phi_3\phi_4\op{O}}^{nm}
	\Braket{\frac{a_1,a_2,a_3,a_4}{r_1,r_2,r_3,r_4}}_{s;nm}
	g^{\Delta_{12},\Delta_{34}}_{\op{O}}(u,v),
\end{multline}
where
$g^{\Delta_{12},\Delta_{34}}_{\op{O}}(u,v)$ is the conformal block in the notation of \cite{Kos:2014bka}, and
\begin{gather}
	\alpha_{\phi_1\phi_2\phi_3\phi_4\op{O}}^{nm}:=
	\lambda_{\phi_1\phi_2\op{O}}^n\lambda_{\phi_3\phi_4\bar{\op{O}}}^m
	=\dim(s) 	\alpha_{\phi_1\phi_2\op{O}}^n\alpha_{\phi_3\phi_4\bar{\op{O}}}^m,
	\label{alpha1234} \\
x_{ij}=|x_i-x_j|,\quad
\Delta_{ij}=|\Delta_i-\Delta_j|,\quad
u=\frac{x_{12}^2x_{34}^2}{x_{13}^2x_{24}^2},\qquad
v=\frac{x_{14}^2x_{23}^2}{x_{13}^2x_{24}^2}.
\end{gather}

Symmetrizing/anti-symmetrizing in $u$ and $v$, we obtain the bootstrap equation in the form
\begin{equation}
\begin{aligned}
	0&=\sum_s\sum_{\op{O}:s}\sum_{nm}\sigma(\op{O})\alpha_{1234\op{O}}^{nm}
	\Braket{\frac{a_1,a_2,a_3,a_4}{r_1,r_2,r_3,r_4}}_{s;nm}F_{\mp,\op{O}}^{12,34}(u,v)\\
	&\pm\sum_t\sum_{\op{O}:t}\sum_{kl}\sigma(\op{O})\alpha_{1432\op{O}}^{kl}
	\Braket{\frac{a_1,a_4,a_3,a_2}{r_1,r_4,r_3,r_2}}_{t;kl}F_{\mp,\op{O}}^{14,32}(u,v)
\end{aligned}
\end{equation}
where \begin{equation}
F^{ij,kl}_{\mp,\op{O}}(u,v)=
v^{(\Delta_k+\Delta_j)/2} g^{\Delta_{ij},\Delta_{kl}}_{\op{O}}(u,v)
\mp
u^{(\Delta_k+\Delta_j)/2} g^{\Delta_{ij},\Delta_{kl}}_{\op{O}}(v,u).
\end{equation}
We note that this function satisfies the relations\footnote{%
The authors thank Shai Chester for pointing out the importance of removing redundant equations using these relations, in particular \eqref{redundant}.
}\begin{align}
F^{ij,kl}_{\mp,\op{O}}(u,v)&=F^{kl,ij}_{\mp,\op{O}}(u,v),\\
F^{ij,kl}_{\mp,\op{O}}(u,v)&=F^{ji,lk}_{\mp,\op{O}}(u,v),\label{redundant}
\end{align} which follow from the properties of $ g^{\Delta_{ij},\Delta_{kl}}_{\op{O}}(v,u)$, see e.g.~Eq.~(59) of \cite{Poland:2018epd}.
We automatically reorder $i,j,k,l$ by these symmetries during the calculation.

We take the inner product with $\Braket{\frac{a_1,a_2,a_3,a_4}{r_1,r_2,r_3,r_4}}_{s;nm}$.
We find
\begin{equation}
	0=F_{\mp,s;nm}^{1234}(u,v)\pm\sum_{t;kl}\six
	F_{\mp,t;kl}^{1432}(u,v)\label{foo}
\end{equation} where \begin{equation}
	F_{\mp,s;nm}^{1234}(u,v):=\sum_{\op{O}:s}
	\sigma(\op{O})\alpha_{1234\op{O}}^{nm}F_{\mp,\op{O}}^{12,34}(u,v).
\end{equation}

Let us assume the existence of external primary scalar operators $\phi_i $
in the representation $r_i\in\Irr(G)$ and the scaling dimension $\Delta_i$.
For each choice of four external operators $\phi_{1,2,3,4}$,
the sign $\mp$,
and the intermediate channel $s;nm$,
we have a bootstrap equation of the form \eqref{foo}.
We denote such a choice by $C$.
For each choice $C$, the equation involves a sum over all possible intermediate operators $\op{O}$.
In the equation \eqref{foo} for each choice $C$,
there appear quadratic combinations \eqref{alpha1234} of possibly complex numbers $\alpha_{ij\op{O}}^n$
which are in turn linear combinations of real numbers $\beta_{ij\op{O}}^m$
introduced at the end of Sec.~\ref{sec:OPE}.
For a given intermediate operator $\op{O}$,
we now uniformly write all of $\beta_{ij\op{O}}^m$ appearing in the equations \eqref{foo}
obtained by varying the choice $C$ by $\beta_{I\op{O}}$ with $I=1,2,\ldots$.
Then the entire set of bootstrap equations has the form \begin{equation}
\sum_{\op{O}}\sum_{I,J} \beta_{I\op{O}} \beta_{J\op{O}} F^{IJ,C}_{\op{O}}(u,v) =0.
\label{FIJ}
\end{equation}

We now denote the space of functions on $(u,v)$ by $\mathcal{F}$ and
consider a vector of functionals $f_C:\mathcal{F}\to \mathbb{R}$, indexed by the choice $C$.
We can exclude the existence of such a CFT
if we can find $f_C$ such that \begin{equation}
\sum_C f_C\cdot F^{IJ,C}_{\op{O}} \succ 0,
\end{equation}
where $A^{IJ}\succ 0$ means that $A$ is a positive-definite matrix.

In practice, we classify the intermediate operators into \emph{sectors},
specified by either the identity $1$, or known external operators $\phi_i$, or unknown operators specified by $r\in\Irr(G)$ and the spin $l$.
We then demand that \begin{equation}
\sum_C f_C\cdot F^{IJ,C}_{1} \succeq 0
\end{equation} in the identity sector, \begin{equation}
\sum_C f_C\cdot F^{IJ,C}_{\phi_i} \succ 0
\end{equation} for each external operator $\phi_i$ with an assumed dimension $\Delta_i$,
and \begin{equation}
\sum_C f_C\cdot F^{IJ,C}_{\op{O}} \succ 0
\end{equation} for other intermediate primary operators $\op{O}$ in the representation $r$ and the spin $l$ with a specified gap condition.
When $l>0$ we usually simply impose the unitarity bound $\Delta(\op{O})\ge l+(d-2)/2$.
For the scalars the gap condition depends on the physics constraints one wants to impose on the spectrum of the theory.

\section{Implementation}
\label{sec:implementation}
\subsection{Group theory data}
In \code{autoboot} we provide a proof-of-concept implementation of the strategy described in the previous section.
For each compact group $G$ to be supported in \code{autoboot},
one needs to provide the following information:
\begin{itemize}
\item Labels $r$ of irreducible representations together with their dimensions
\item The complex conjugation map $r \mapsto \bar r$.
\item Abstract tensor product decompositions of $r_i\otimes r_j$ into irreducible representations
\item Explicit unitary representation matrices of the generators of $G$ for each irreducible representation $r$.
\end{itemize}

Currently we support 
small finite groups $G$ in the \code{SmallGrp} library \cite{SmallGrp} of the computer algebra system \code{GAP} \cite{GAP}
and small classical groups $G=SO(2),O(2),SO(3),O(3),U(1),SU(2)$.
For classical groups, these data can in principle be generated automatically,
but at present we implement by hand only a few representations we actually support.

For small finite groups, we use a separate script to extract these data from \code{GAP} and convert them using a \code{C\#} program into a form easily usable from \code{autoboot}.
Currently the script uses \code{IrreducibleRepresentationsDixon} in the \code{GAP} library \code{ctbllib},
which is based on the algorithm described in \cite{Dixon}.
Due to the slowness of this algorithm, the distribution of \code{autoboot} as of March 2019 does not contain the converted data for all the small groups in the \code{SmallGrp} library.
If any reader needs to generate the data for a small group not contained in the distribution of \code{autoboot}, please ask the authors for assistance.
A faster  function 
to generate irreducible representations is available in the \code{GAP} library \code{repsn} \cite{REPSN} based on the paper \cite{REPSNpaper}, but unfortunately it does not give unitary matrices at present.

We have in fact implemented two variants, one where matrix elements are computed as algebraic numbers, and another where matrix elements are numerically evaluated.
The line
\begin{lstlisting}
	<<"group.m"
\end{lstlisting}
or 
\begin{lstlisting}
	<<"ngroup.m"
\end{lstlisting}
loads the algebraic or numerical version, respectively.

The invariant tensor
$f(a,b,c)=\code{ope[$r,s,t$][$n$][$a,b,c$]}$ needs to satisfy
\begin{equation}
	\sum_{a'b'}r(g)_{aa'}s(g)_{bb'}f(a',b',c)
	=\sum_{c'}f(a,b,c')t(g)_{c'c}
\end{equation} for the discrete part $g\in G$
and
\begin{equation}
	\sum_{a'}r(x)_{aa'}f(a',b,c)
	+\sum_{b'}s(x)_{bb'}f(a,b',c)
	=\sum_{c'}f(a,b,c')t(x)_{c'c}
\end{equation}
for infinitesimal generators $x\in \mathfrak{g}$,
where we use $r_{aa'}$ for the representation matrices for a representation $r$, etc.
Our \code{autoboot} enumerates these equations from the given explicit representation matrices, and solves them using \code{NullSpace} and \code{Orthogonalize} of \code{Mathematica}.
We also make sure that for $r=s$ these coefficients are either even or odd under $a\leftrightarrow b$.

The notations in this paper and in the code are mapped as follows:
\begin{align}
	\code{inv[$r,s,t$]}&=\dim\inv{r,s,t},\\
	\code{ope[$r,s,t$][$n$][$a,b,c$]}&=\Set{\frac{c}{t}|\frac{a,b}{r,s}}_n,\\
	\code{ope[$r$][$a,b$]}&=\Set{a,b}_r,\\
	\code{cor[$r,s,t$][$n$][$a,b,c$]}&=\Braket{\frac{a,b,c}{r,s,t}}_n,\\
	\code{cor[$r_1,r_2,r_3,r_4$][$s,n,m$][$a_1,a_2,a_3,a_4$]}
	&=\Braket{\frac{a_1,a_2,a_3,a_4}{r_1,r_2,r_3,r_4}}_{s;nm}.
\end{align}
Various isomorphisms among the invariant tensors are given by the following:
\begin{align}
	\code{$\sigma$[$r,s,t$][$n$]}&=\sigma_n(r,s,t),\\
	\code{$\tau$[$r,s,t$][$n,m$]}&=\tau_{nm}(r,s,t),\\
	\code{$\omega$[$r,s,t$][$n,m$]}&=\omega_{nm}(r,s,t),\\
	\code{six[$r_1,r_2,r_3,r_4$][$s,n,m,t,k,l$]}&=\six.
\end{align}
These (except $\sigma$) can be computed using the inner product of the invariant tensors as explained already.
Since the matrix elements of invariant tensors are often very sparse, and that the dimension of the space of invariant tensors is often simply 1, our \code{autoboot} uses a quicker method in computing them, by using only the first few nonzero entries of the invariant tensors and actually solving the linear equations.

\subsection{CFT data}
A primary operator $\op O$ in the representation $r$ in $G$,
with the sign $\sigma(\op{O})=p$ and the spin $l$ such that $(-1)^l=q$
is represented by
\begin{equation}
	\op{O}=\code{op[$\op{O}$,$r$,$p$,$q$]}.
\end{equation}
The complex conjugate operator is then \code{dualOp[$\op{O}$]}.
The first argument is the name of the operator;
all intermediate operators share the name \code{op}.
The unit operator is given by $1=\code{op[0,$\id$,1,1]}$.
To register an external primary scalar operator, call
\begin{equation}
\code{setOps[\{op[x,r,p,q],...\}]}.
\end{equation}
As a shorthand, we can use \code{op[x,r]} for \code{op[x,r,1,1]}.

The OPE coefficients are denoted by
\begin{align}
	\code{$\lambda$[$\phi_1,\phi_2,\op{O}$][$n$]}&=\lambda_{\phi_1\phi_2\op{O}}^n, &
	\code{$\alpha$[$\phi_1,\phi_2,\op{O}$][$n$]}&=\alpha_{\phi_1\phi_2\op{O}}^n
\end{align}
for intermediate operators $\op{O}$, and by
\begin{align}
	\code{$\mu$[$\phi_1,\phi_2,\phi_3$][$n$]}&=\lambda_{\phi_1\phi_2\phi_3}^n,&
	\code{$\nu$[$\phi_1,\phi_2,\phi_3$][$n$]}&=\alpha_{\phi_1\phi_2\phi_3}^n
\end{align}
for external operators $\phi_3=\op{O}$.
Internally, we solve the constraints \eqref{eq:ope12}, \eqref{eq:opetau}, \eqref{eq:opeomega}
as explained at the end of Sec.~\ref{sec:OPE}
and represent them all by linear combinations of real constants \code{$\beta$[$\phi_1,\phi_2,\op{O}$][$m$]}.

\subsection{Bootstrap equations}
The bootstrap equations are obtained by calling \code{bootAll[]}.
When the bootstrap equations are given by $a=0\land b=0\land\cdots$,
the return value of \code{bootAll[]} is \code{eqn[\{$a,b,\ldots$\}]}.
Here, $a,b,\ldots$ are given by real linear combinations of \code{sum} and \code{single},
where \code{sum[f,op[x,$r$,$p$,$q$]]} represents
\begin{equation}
	\sum_{\substack{\op{O}:r\\\sigma(\op{O})=p\\(-1)^l=q}}f
\end{equation}
and \code{single[f]} corresponds to just $f$.

Inside the code, the conformal blocks are represented by
\begin{align}
	\code{Fp[a,b,c,d,o]}&=F_{-,o}^{ab,cd},&
	\code{Hp[a,b,c,d,o]}&=F_{+,o}^{ab,cd},\\
	\code{F[a,b,c,d]}&=F_{-,\op{O}}^{ab,cd},&
	\code{H[a,b,c,d]}&=F_{+,\op{O}}^{ab,cd}.
\end{align}
Then the function $f$ inside \code{single} is a product of two \code{$\beta$}'s
and \code{Fp} or \code{Hp},
and the function $f$ inside \code{sum} is a product of two \code{$\beta$}'s
and \code{F} or \code{H}.
The function \code{format} gives a more readable representation of the equations.

We convert the bootstrap equations into a semi-definite program following the standard method.
To do this, \code{makeSDP[...]} first finds all the sectors $\op{O}$ in the intermediate channel,
and for each sector, we list all the OPE coefficients $\beta_{I\op{O}}$ involved in that sector.
We then extract the vector of matrices $F^{IJ,C}$ as described in \eqref{FIJ}.
In practice, this matrix $F^{IJ,C}$ is very sparse and automatically block-diagonal;
 \code{autoboot} splits it up accordingly.

At this point, it is straightforward to convert it into a form understandable by a semi-definite program solver \code{spdb}.
We implemented a function \code{toCboot[...]} which constructs a \code{Python} program which uses \code{cboot} \cite{Nakayama:2016jhq}.
After saving the \code{Python} program to a file, some minor edits will be necessary
to set up the gaps in the assumed spectrum etc.
Then the \code{Python} program will output the file which can be fed to \code{sdpb}.

\section{Examples}
\label{sec:examples}
\subsection{3d Ising model with $\epsilon$ and $\sigma$}
Let us now reproduce the ground-breaking result of \cite{Kos:2014bka}, where the mixed-correlator bootstrap of the 3d Ising model with the energy operator $\epsilon$ and the spin operator $\sigma$ was first performed.
We use the following code
\begin{lstlisting}
	z2=getGroup[2,1];
	setGroup[z2];
	setOps[{op[e,z2[id]], op[a,rep[2]]}];
	eq=bootAll[];
	sdp=makeSDP[eq];
	py=toCboot[sdp];
	WriteString["Ising.py",py];
\end{lstlisting}
Here, we set \code{z2} to the first group with two elements, namely $\mathbb{Z}_2$.
We then introduce a $\mathbb{Z}_2$-even operator $\epsilon=\code{e}$
and a $\mathbb{Z}_2$-odd operator $\sigma=\code{a}$.
We use the symbol \code{a} to represent the operator $\sigma$, since the symbol given to \code{op} will also be used in the generated \code{Python} code.
Here we also illustrated another small feature of \code{autoboot}, where the trivial representation for a group $g$ can be found by \code{$g$[id]}.
We save the \code{Python} code into \code{Ising.py}.

The resulting \code{Python} code uses \code{cboot}.
We need to make a few manual modifications to the first few lines in the file:
\begin{lstlisting}[language=Python]
	context=cb.context_for_scalar(epsilon=0.5,Lambda=11)
	spins=list(range(22))
	nu_max=8
	mygap={}
\end{lstlisting}
Let us explain it line by line: \begin{itemize}
\item $\code{epsilon}=(d-2)/2$ is given by the spacetime dimension $d$,
\item $\code{Lambda}=\Lambda$ specifies the cutoff $m+n\le \Lambda$ in the derivative expansion of the conformal blocks $\partial_u^m \partial_v^n F(u,v)$,
\item \code{spins} controls the spins $\ell$ in the intermediate channel to consider,
\item $\code{nu\_max}=\nu_\text{max}$ is the number of poles which will be used in the numerical computation of the conformal block as explained in \cite{Kos:2014bka},
\item and finally \code{mygap} specifies the gap for each unnamed operator in the intermediate channel in the format
\code{\{($r$,$\ell$):$\Delta$, \ldots \}} where $r$ is the dimension, $\ell$ is the spin, and $\Delta$ is the lowest allowed scaling dimension.
If not explicitly specified, the code assumes the unitarity bound.
For example, in our case, we can set it to
\begin{lstlisting}[language=Python]
	mygap={("rep[1]",0): 3, ("rep[2]",0): 3}
\end{lstlisting}
to assume that all unnamed scalar operators are irrelevant.
\end{itemize}

To actually create an input to \code{sdpb}, we perform
\begin{lstlisting}[language=Python]
	write_SDP({"e": 1.4127, "a": 0.5181})
\end{lstlisting}
which will generate the semi-definite program for a given $\Delta_\epsilon$ and $\Delta_\sigma$.
We can run the resulting program and check that the output is consistent with the results of \cite{Kos:2014bka,Simmons-Duffin:2015qma}.
We can further enhance the program to search for a certain region in the $(\Delta_\epsilon,\Delta_\sigma)$ space,
and/or to run \code{sdpb} from within the program, and so on.

\code{autoboot} also has the ability to generate the bootstrap equations in \LaTeX, to be used enclosed in the \verb+\begin{align}...\end{align}+ environment.
To use this facility, we set up the mapping between the \code{Mathematica} names of the operators and  the representations and their \LaTeX\ counterpart, and call \code{toTeX}.
In the case of the 3d Ising model, we do for example
\begin{lstlisting}
	opToTeX[e] := "\\epsilon"
	opToTeX[a] := "\\sigma"
	repToTeX[rep[1]] := "I^+"
	repToTeX[rep[2]] := "I^-"
	toTeX[eq]
\end{lstlisting} which generates the following set of bootstrap equations:

{\footnotesize
\begin{align*}
0&=F^{\sigma \sigma,\sigma \sigma}_{-,1}+{\lambda_{\sigma \sigma \epsilon}}^2 F^{\sigma \sigma,\sigma \sigma}_{-,\epsilon}+\sum_{\op{O}^+:{I^+}}{\lambda_{\sigma \sigma \op{O}}}^2 F^{\sigma \sigma,\sigma \sigma}_{-,\op{O}},\\
0&=F^{\epsilon \epsilon,\epsilon \epsilon}_{-,1}+{\lambda_{\epsilon \epsilon \epsilon}}^2 F^{\epsilon \epsilon,\epsilon \epsilon}_{-,\epsilon}+\sum_{\op{O}^+:{I^+}}{\lambda_{\epsilon \epsilon \op{O}}}^2 F^{\epsilon \epsilon,\epsilon \epsilon}_{-,\op{O}},\\
0&={\lambda_{\sigma \sigma \epsilon}}^2 F^{\epsilon \sigma,\epsilon \sigma}_{-,\sigma}+\sum_{\op{O}^-:{I^-}}{\lambda_{\epsilon \sigma \op{O}}}^2 F^{\epsilon \sigma,\epsilon \sigma}_{-,\op{O}}+\sum_{\op{O}^+:{I^-}}{\lambda_{\epsilon \sigma \op{O}}}^2 F^{\epsilon \sigma,\epsilon \sigma}_{-,\op{O}},\\
0&=F^{\epsilon \epsilon,\sigma \sigma}_{-,1}+{\lambda_{\sigma \sigma \epsilon}}^2 F^{\epsilon \sigma,\sigma \epsilon}_{-,\sigma}+\lambda_{\sigma \sigma \epsilon} \lambda_{\epsilon \epsilon \epsilon} F^{\epsilon \epsilon,\sigma \sigma}_{-,\epsilon}-\sum_{\op{O}^-:{I^-}}{\lambda_{\epsilon \sigma \op{O}}}^2 F^{\epsilon \sigma,\sigma \epsilon}_{-,\op{O}}+\sum_{\op{O}^+:{I^-}}{\lambda_{\epsilon \sigma \op{O}}}^2 F^{\epsilon \sigma,\sigma \epsilon}_{-,\op{O}}+\sum_{\op{O}^+:{I^+}}\lambda_{\sigma \sigma \op{O}} \lambda_{\epsilon \epsilon \op{O}} F^{\epsilon \epsilon,\sigma \sigma}_{-,\op{O}},\\
0&=F^{\epsilon \epsilon,\sigma \sigma}_{+,1}-{\lambda_{\sigma \sigma \epsilon}}^2 F^{\epsilon \sigma,\sigma \epsilon}_{+,\sigma}+\lambda_{\sigma \sigma \epsilon} \lambda_{\epsilon \epsilon \epsilon} F^{\epsilon \epsilon,\sigma \sigma}_{+,\epsilon}+\sum_{\op{O}^-:{I^-}}{\lambda_{\epsilon \sigma \op{O}}}^2 F^{\epsilon \sigma,\sigma \epsilon}_{+,\op{O}}-\sum_{\op{O}^+:{I^-}}{\lambda_{\epsilon \sigma \op{O}}}^2 F^{\epsilon \sigma,\sigma \epsilon}_{+,\op{O}}+\sum_{\op{O}^+:{I^+}}\lambda_{\sigma \sigma \op{O}} \lambda_{\epsilon \epsilon \op{O}} F^{\epsilon \epsilon,\sigma \sigma}_{+,\op{O}}.
\end{align*}
}

\subsection{3d $O(2)$ model with three external scalar operators}
As the next example, we consider the 3d $O(2)$ model
with three primary scalar external operators: a singlet $s$, a vector $\phi$ and a traceless-symmetric $t$.
In \cite{Kos:2015mba}, the same model was analyzed using $s$ and $\phi$ as external operators,
and the information on $\Delta_t$ was obtained by specifying the condition in the intermediate channel.
The \code{Mathematica} code required is simply:
\begin{lstlisting}
	o2=getO[2];
	setGroup[o2];
	setOps[{op[s,o2[id]], op[v,v[1]], op[t,v[2]]}];
	eq=bootAll[];
	sdp=makeSDP[eq];
	py=toCboot[sdp];
	WriteString["O2.py",py];
\end{lstlisting}
Here \code{o2=getO[2]} creates the group $O(2)$ within \code{autoboot},
and \code{v[n]} stands for the \code{n}-th traceless symmetric representation.
We use the symbol \code{v} to represent the operator $\phi$.
The  bootstrap equations generated by \code{toTeX} are given below:

{\tiny\allowdisplaybreaks
\begin{align*}
	0&=F^{s s,s s}_{-,1}+{\lambda_{s s s}}^2 F^{s s,s s}_{-,s}+\sum_{\op{O}^+:{\mathbf{I}^+}}{\lambda_{s s \op{O}}}^2 F^{s s,s s}_{-,\op{O}},\\
	0&=\sum_{\op{O}^-:{\mathbf{I}^-}}{\lambda_{t t \op{O}}}^2 F^{t t,t t}_{-,\op{O}}+\frac{1}{2}\sum_{\op{O}^+:{\mathbf{S}^{4}}}{\lambda_{t t \op{O}}}^2 F^{t t,t t}_{-,\op{O}},\\
	0&={\lambda_{t t s}}^2 F^{s t,s t}_{-,t}+\sum_{\op{O}^-:{\mathbf{T}}}{\lambda_{s t \op{O}}}^2 F^{s t,s t}_{-,\op{O}}+\sum_{\op{O}^+:{\mathbf{T}}}{\lambda_{s t \op{O}}}^2 F^{s t,s t}_{-,\op{O}},\\
	0&={\lambda_{\phi \phi s}}^2 F^{s \phi,s \phi}_{-,\phi}+\sum_{\op{O}^-:{\mathbf{V}}}{\lambda_{s \phi \op{O}}}^2 F^{s \phi,s \phi}_{-,\op{O}}+\sum_{\op{O}^+:{\mathbf{V}}}{\lambda_{s \phi \op{O}}}^2 F^{s \phi,s \phi}_{-,\op{O}},\\
	0&=F^{t t,t t}_{-,1}+\frac{1}{2}{\lambda_{t t s}}^2 F^{t t,t t}_{-,s}+\frac{1}{2}\sum_{\op{O}^+:{\mathbf{I}^+}}{\lambda_{t t \op{O}}}^2 F^{t t,t t}_{-,\op{O}}+\frac{1}{4}\sum_{\op{O}^+:{\mathbf{S}^{4}}}{\lambda_{t t \op{O}}}^2 F^{t t,t t}_{-,\op{O}},\\
	0&={\lambda_{\phi t \phi}}^2 F^{\phi \phi,\phi \phi}_{-,t}+2\sum_{\op{O}^-:{\mathbf{I}^-}}{\lambda_{\phi \phi \op{O}}}^2 F^{\phi \phi,\phi \phi}_{-,\op{O}}+\sum_{\op{O}^+:{\mathbf{T}}}{\lambda_{\phi \phi \op{O}}}^2 F^{\phi \phi,\phi \phi}_{-,\op{O}},\\
	0&=\lambda_{\phi t \phi} \lambda_{\phi \phi s} F^{s \phi,t \phi}_{-,\phi}-\sum_{\op{O}^-:{\mathbf{V}}}\lambda_{s \phi \op{O}} \lambda_{\phi t \op{O}} F^{s \phi,t \phi}_{-,\op{O}}+\sum_{\op{O}^+:{\mathbf{V}}}\lambda_{s \phi \op{O}} \lambda_{\phi t \op{O}} F^{s \phi,t \phi}_{-,\op{O}},\\
	0&=F^{\phi \phi,\phi \phi}_{-,1}+\frac{1}{2}{\lambda_{\phi \phi s}}^2 F^{\phi \phi,\phi \phi}_{-,s}-\frac{1}{2}\sum_{\op{O}^-:{\mathbf{I}^-}}{\lambda_{\phi \phi \op{O}}}^2 F^{\phi \phi,\phi \phi}_{-,\op{O}}+\frac{1}{2}\sum_{\op{O}^+:{\mathbf{I}^+}}{\lambda_{\phi \phi \op{O}}}^2 F^{\phi \phi,\phi \phi}_{-,\op{O}},\\
	0&=F^{t t,t t}_{+,1}+\frac{1}{2}{\lambda_{t t s}}^2 F^{t t,t t}_{+,s}+\frac{1}{2}\sum_{\op{O}^-:{\mathbf{I}^-}}{\lambda_{t t \op{O}}}^2 F^{t t,t t}_{+,\op{O}}+\frac{1}{2}\sum_{\op{O}^+:{\mathbf{I}^+}}{\lambda_{t t \op{O}}}^2 F^{t t,t t}_{+,\op{O}}-\frac{1}{2}\sum_{\op{O}^+:{\mathbf{S}^{4}}}{\lambda_{t t \op{O}}}^2 F^{t t,t t}_{+,\op{O}},\\
	0&={\lambda_{\phi t \phi}}^2 F^{\phi t,\phi t}_{-,\phi}+\sum_{\op{O}^-:{\mathbf{V}}}{\lambda_{\phi t \op{O}}}^2 F^{\phi t,\phi t}_{-,\op{O}}+\sum_{\op{O}^+:{\mathbf{V}}}{\lambda_{\phi t \op{O}}}^2 F^{\phi t,\phi t}_{-,\op{O}}+\sum_{\op{O}^-:{\mathbf{S}^{3}}}{\lambda_{\phi t \op{O}}}^2 F^{\phi t,\phi t}_{-,\op{O}}+\sum_{\op{O}^+:{\mathbf{S}^{3}}}{\lambda_{\phi t \op{O}}}^2 F^{\phi t,\phi t}_{-,\op{O}},\\
	0&={\lambda_{\phi t \phi}}^2 F^{\phi t,\phi t}_{+,\phi}+\sum_{\op{O}^-:{\mathbf{V}}}{\lambda_{\phi t \op{O}}}^2 F^{\phi t,\phi t}_{+,\op{O}}+\sum_{\op{O}^+:{\mathbf{V}}}{\lambda_{\phi t \op{O}}}^2 F^{\phi t,\phi t}_{+,\op{O}}-\sum_{\op{O}^-:{\mathbf{S}^{3}}}{\lambda_{\phi t \op{O}}}^2 F^{\phi t,\phi t}_{+,\op{O}}-\sum_{\op{O}^+:{\mathbf{S}^{3}}}{\lambda_{\phi t \op{O}}}^2 F^{\phi t,\phi t}_{+,\op{O}},\\
	0&=F^{s s,t t}_{-,1}+\frac{\sqrt{2}}{2}\lambda_{s s s} \lambda_{t t s} F^{s s,t t}_{-,s}+\frac{1}{2}{\lambda_{t t s}}^2 F^{s t,t s}_{-,t}-\frac{1}{2}\sum_{\op{O}^-:{\mathbf{T}}}{\lambda_{s t \op{O}}}^2 F^{s t,t s}_{-,\op{O}}+\frac{1}{2}\sum_{\op{O}^+:{\mathbf{T}}}{\lambda_{s t \op{O}}}^2 F^{s t,t s}_{-,\op{O}}+\frac{\sqrt{2}}{2}\sum_{\op{O}^+:{\mathbf{I}^+}}\lambda_{s s \op{O}} \lambda_{t t \op{O}} F^{s s,t t}_{-,\op{O}},\\
	0&=F^{s s,t t}_{+,1}+\frac{\sqrt{2}}{2}\lambda_{s s s} \lambda_{t t s} F^{s s,t t}_{+,s}-\frac{1}{2}{\lambda_{t t s}}^2 F^{s t,t s}_{+,t}+\frac{1}{2}\sum_{\op{O}^-:{\mathbf{T}}}{\lambda_{s t \op{O}}}^2 F^{s t,t s}_{+,\op{O}}-\frac{1}{2}\sum_{\op{O}^+:{\mathbf{T}}}{\lambda_{s t \op{O}}}^2 F^{s t,t s}_{+,\op{O}}+\frac{\sqrt{2}}{2}\sum_{\op{O}^+:{\mathbf{I}^+}}\lambda_{s s \op{O}} \lambda_{t t \op{O}} F^{s s,t t}_{+,\op{O}},\\
	0&=\lambda_{t t s} \lambda_{\phi t \phi} F^{s t,\phi \phi}_{-,t}+\lambda_{\phi t \phi} \lambda_{\phi \phi s} F^{s \phi,\phi t}_{-,\phi}+\sum_{\op{O}^-:{\mathbf{V}}}\lambda_{s \phi \op{O}} \lambda_{\phi t \op{O}} F^{s \phi,\phi t}_{-,\op{O}}+\sum_{\op{O}^+:{\mathbf{V}}}\lambda_{s \phi \op{O}} \lambda_{\phi t \op{O}} F^{s \phi,\phi t}_{-,\op{O}}-\sum_{\op{O}^+:{\mathbf{T}}}\lambda_{s t \op{O}} \lambda_{\phi \phi \op{O}} F^{s t,\phi \phi}_{-,\op{O}},\\
	0&=\lambda_{t t s} \lambda_{\phi t \phi} F^{s t,\phi \phi}_{+,t}-\lambda_{\phi t \phi} \lambda_{\phi \phi s} F^{s \phi,\phi t}_{+,\phi}-\sum_{\op{O}^-:{\mathbf{V}}}\lambda_{s \phi \op{O}} \lambda_{\phi t \op{O}} F^{s \phi,\phi t}_{+,\op{O}}-\sum_{\op{O}^+:{\mathbf{V}}}\lambda_{s \phi \op{O}} \lambda_{\phi t \op{O}} F^{s \phi,\phi t}_{+,\op{O}}-\sum_{\op{O}^+:{\mathbf{T}}}\lambda_{s t \op{O}} \lambda_{\phi \phi \op{O}} F^{s t,\phi \phi}_{+,\op{O}},\\
	0&=F^{\phi \phi,\phi \phi}_{+,1}-\frac{1}{2}{\lambda_{\phi t \phi}}^2 F^{\phi \phi,\phi \phi}_{+,t}+\frac{1}{2}{\lambda_{\phi \phi s}}^2 F^{\phi \phi,\phi \phi}_{+,s}+\frac{1}{2}\sum_{\op{O}^-:{\mathbf{I}^-}}{\lambda_{\phi \phi \op{O}}}^2 F^{\phi \phi,\phi \phi}_{+,\op{O}}+\frac{1}{2}\sum_{\op{O}^+:{\mathbf{I}^+}}{\lambda_{\phi \phi \op{O}}}^2 F^{\phi \phi,\phi \phi}_{+,\op{O}}-\frac{1}{2}\sum_{\op{O}^+:{\mathbf{T}}}{\lambda_{\phi \phi \op{O}}}^2 F^{\phi \phi,\phi \phi}_{+,\op{O}},\\
	0&=F^{s s,\phi \phi}_{-,1}+\frac{\sqrt{2}}{2}\lambda_{s s s} \lambda_{\phi \phi s} F^{s s,\phi \phi}_{-,s}+\frac{1}{2}{\lambda_{\phi \phi s}}^2 F^{s \phi,\phi s}_{-,\phi}-\frac{1}{2}\sum_{\op{O}^-:{\mathbf{V}}}{\lambda_{s \phi \op{O}}}^2 F^{s \phi,\phi s}_{-,\op{O}}+\frac{1}{2}\sum_{\op{O}^+:{\mathbf{V}}}{\lambda_{s \phi \op{O}}}^2 F^{s \phi,\phi s}_{-,\op{O}}+\frac{\sqrt{2}}{2}\sum_{\op{O}^+:{\mathbf{I}^+}}\lambda_{s s \op{O}} \lambda_{\phi \phi \op{O}} F^{s s,\phi \phi}_{-,\op{O}},\\
	0&=F^{s s,\phi \phi}_{+,1}+\frac{\sqrt{2}}{2}\lambda_{s s s} \lambda_{\phi \phi s} F^{s s,\phi \phi}_{+,s}-\frac{1}{2}{\lambda_{\phi \phi s}}^2 F^{s \phi,\phi s}_{+,\phi}+\frac{1}{2}\sum_{\op{O}^-:{\mathbf{V}}}{\lambda_{s \phi \op{O}}}^2 F^{s \phi,\phi s}_{+,\op{O}}-\frac{1}{2}\sum_{\op{O}^+:{\mathbf{V}}}{\lambda_{s \phi \op{O}}}^2 F^{s \phi,\phi s}_{+,\op{O}}+\frac{\sqrt{2}}{2}\sum_{\op{O}^+:{\mathbf{I}^+}}\lambda_{s s \op{O}} \lambda_{\phi \phi \op{O}} F^{s s,\phi \phi}_{+,\op{O}},\\
	0&={\lambda_{\phi t \phi}}^2 F^{\phi t,t \phi}_{-,\phi}-\sum_{\op{O}^-:{\mathbf{V}}}{\lambda_{\phi t \op{O}}}^2 F^{\phi t,t \phi}_{-,\op{O}}+\sum_{\op{O}^+:{\mathbf{V}}}{\lambda_{\phi t \op{O}}}^2 F^{\phi t,t \phi}_{-,\op{O}}+\sum_{\op{O}^-:{\mathbf{S}^{3}}}{\lambda_{\phi t \op{O}}}^2 F^{\phi t,t \phi}_{-,\op{O}}-\sum_{\op{O}^+:{\mathbf{S}^{3}}}{\lambda_{\phi t \op{O}}}^2 F^{\phi t,t \phi}_{-,\op{O}}-2\sum_{\op{O}^-:{\mathbf{I}^-}}\lambda_{t t \op{O}} \lambda_{\phi \phi \op{O}} F^{\phi \phi,t t}_{-,\op{O}},\\
	0&={\lambda_{\phi t \phi}}^2 F^{\phi t,t \phi}_{+,\phi}-\sum_{\op{O}^-:{\mathbf{V}}}{\lambda_{\phi t \op{O}}}^2 F^{\phi t,t \phi}_{+,\op{O}}+\sum_{\op{O}^+:{\mathbf{V}}}{\lambda_{\phi t \op{O}}}^2 F^{\phi t,t \phi}_{+,\op{O}}+\sum_{\op{O}^-:{\mathbf{S}^{3}}}{\lambda_{\phi t \op{O}}}^2 F^{\phi t,t \phi}_{+,\op{O}}-\sum_{\op{O}^+:{\mathbf{S}^{3}}}{\lambda_{\phi t \op{O}}}^2 F^{\phi t,t \phi}_{+,\op{O}}+2\sum_{\op{O}^-:{\mathbf{I}^-}}\lambda_{t t \op{O}} \lambda_{\phi \phi \op{O}} F^{\phi \phi,t t}_{+,\op{O}},\\
	0&=F^{\phi \phi,t t}_{-,1}+\frac{1}{2}\lambda_{t t s} \lambda_{\phi \phi s} F^{\phi \phi,t t}_{-,s}-\frac{1}{2}\sum_{\op{O}^-:{\mathbf{S}^{3}}}{\lambda_{\phi t \op{O}}}^2 F^{\phi t,t \phi}_{-,\op{O}}+\frac{1}{2}\sum_{\op{O}^+:{\mathbf{S}^{3}}}{\lambda_{\phi t \op{O}}}^2 F^{\phi t,t \phi}_{-,\op{O}}+\frac{1}{2}\sum_{\op{O}^-:{\mathbf{I}^-}}\lambda_{t t \op{O}} \lambda_{\phi \phi \op{O}} F^{\phi \phi,t t}_{-,\op{O}}+\frac{1}{2}\sum_{\op{O}^+:{\mathbf{I}^+}}\lambda_{t t \op{O}} \lambda_{\phi \phi \op{O}} F^{\phi \phi,t t}_{-,\op{O}},\\
	0&=F^{\phi \phi,t t}_{+,1}+\frac{1}{2}\lambda_{t t s} \lambda_{\phi \phi s} F^{\phi \phi,t t}_{+,s}+\frac{1}{2}\sum_{\op{O}^-:{\mathbf{S}^{3}}}{\lambda_{\phi t \op{O}}}^2 F^{\phi t,t \phi}_{+,\op{O}}-\frac{1}{2}\sum_{\op{O}^+:{\mathbf{S}^{3}}}{\lambda_{\phi t \op{O}}}^2 F^{\phi t,t \phi}_{+,\op{O}}+\frac{1}{2}\sum_{\op{O}^-:{\mathbf{I}^-}}\lambda_{t t \op{O}} \lambda_{\phi \phi \op{O}} F^{\phi \phi,t t}_{+,\op{O}}+\frac{1}{2}\sum_{\op{O}^+:{\mathbf{I}^+}}\lambda_{t t \op{O}} \lambda_{\phi \phi \op{O}} F^{\phi \phi,t t}_{+,\op{O}}.
\end{align*}}

Here we used $\mathbf{S}^n$ to denote the $n$-th symmetric traceless tensor representation, which is \code{v[n]} in \code{Mathematica}. 
We also introduced $\mathbf{V}=\mathbf{S}^1$ and $\mathbf{T}=\mathbf{S}^2$ as abbreviations.
$\mathbf{I}^+$ is the trivial representation and $\mathbf{I}^-$ is the sign representation.

This example shows the power of \code{autoboot}.
It is almost trivial to add another external operator using \code{autoboot},
whereas it is quite tedious to work out the form of the bootstrap equations by hand.

\begin{figure}
\centering
\begin{tabular}{c}
	\begin{minipage}{0.50\hsize}
		\centering
		\includegraphics[width=\textwidth]{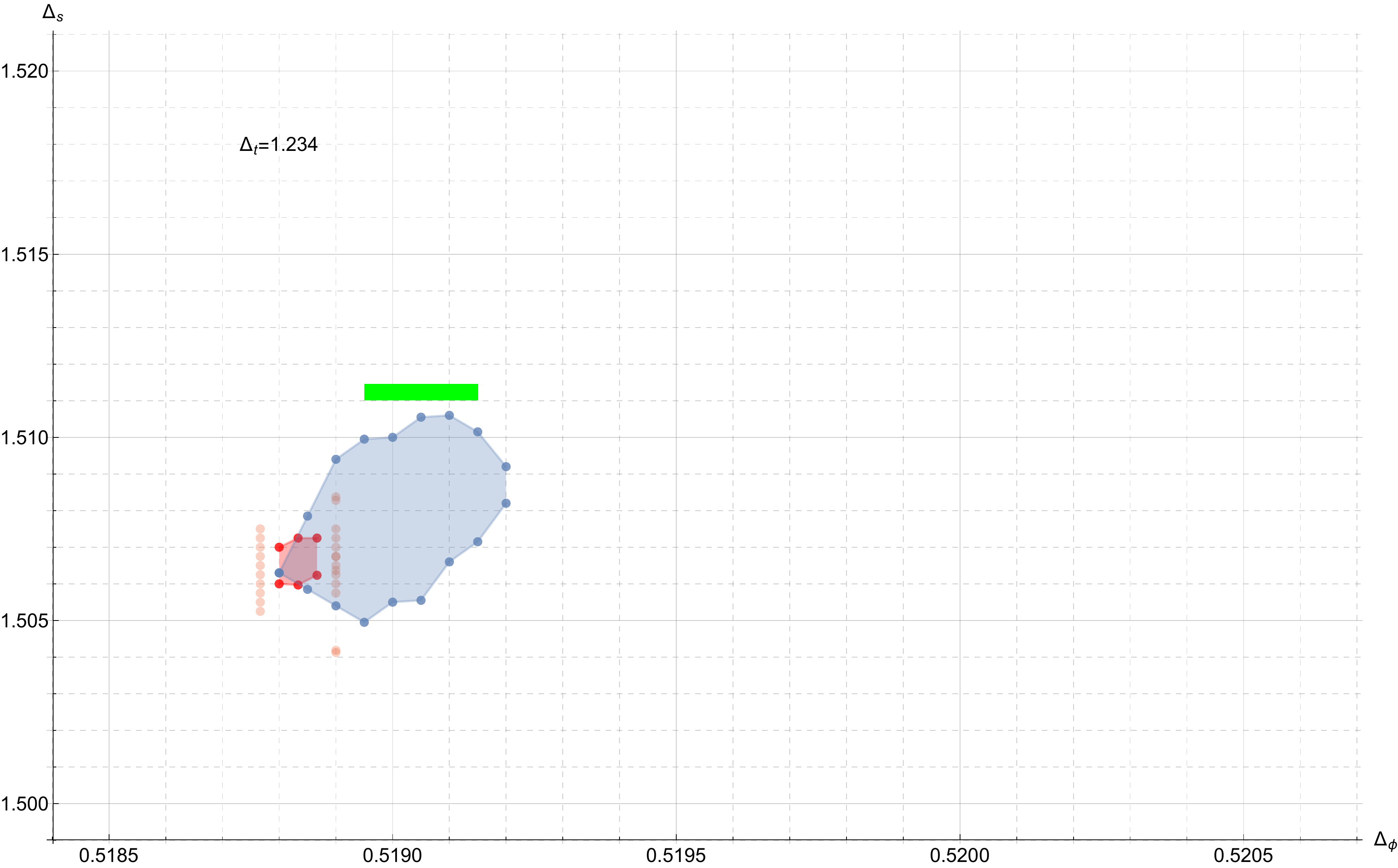}
	\end{minipage}
	\begin{minipage}{0.50\hsize}
		\centering
		\includegraphics[width=\textwidth]{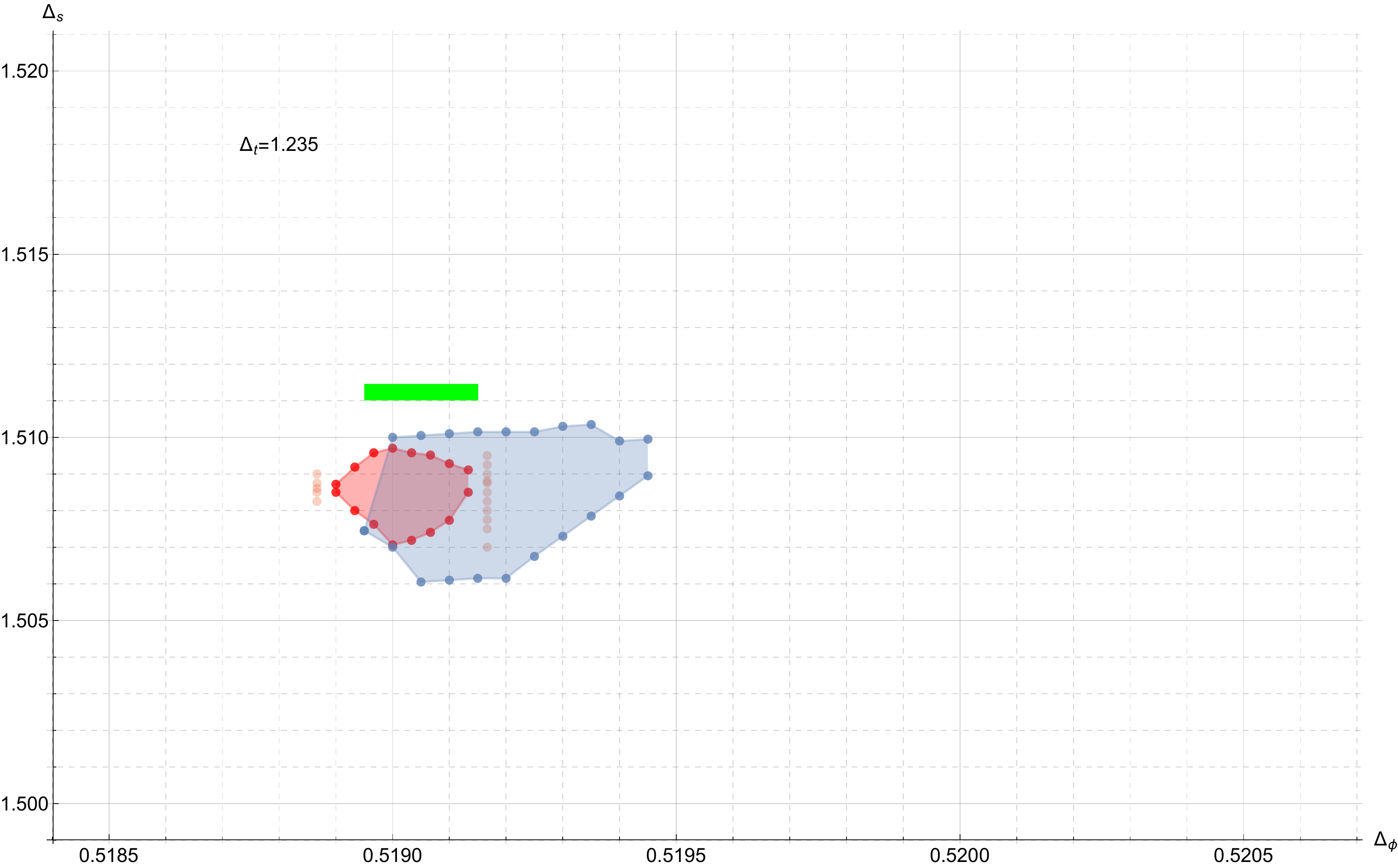}
	\end{minipage}\\
	\begin{minipage}{0.50\hsize}
		\centering
		\includegraphics[width=\textwidth]{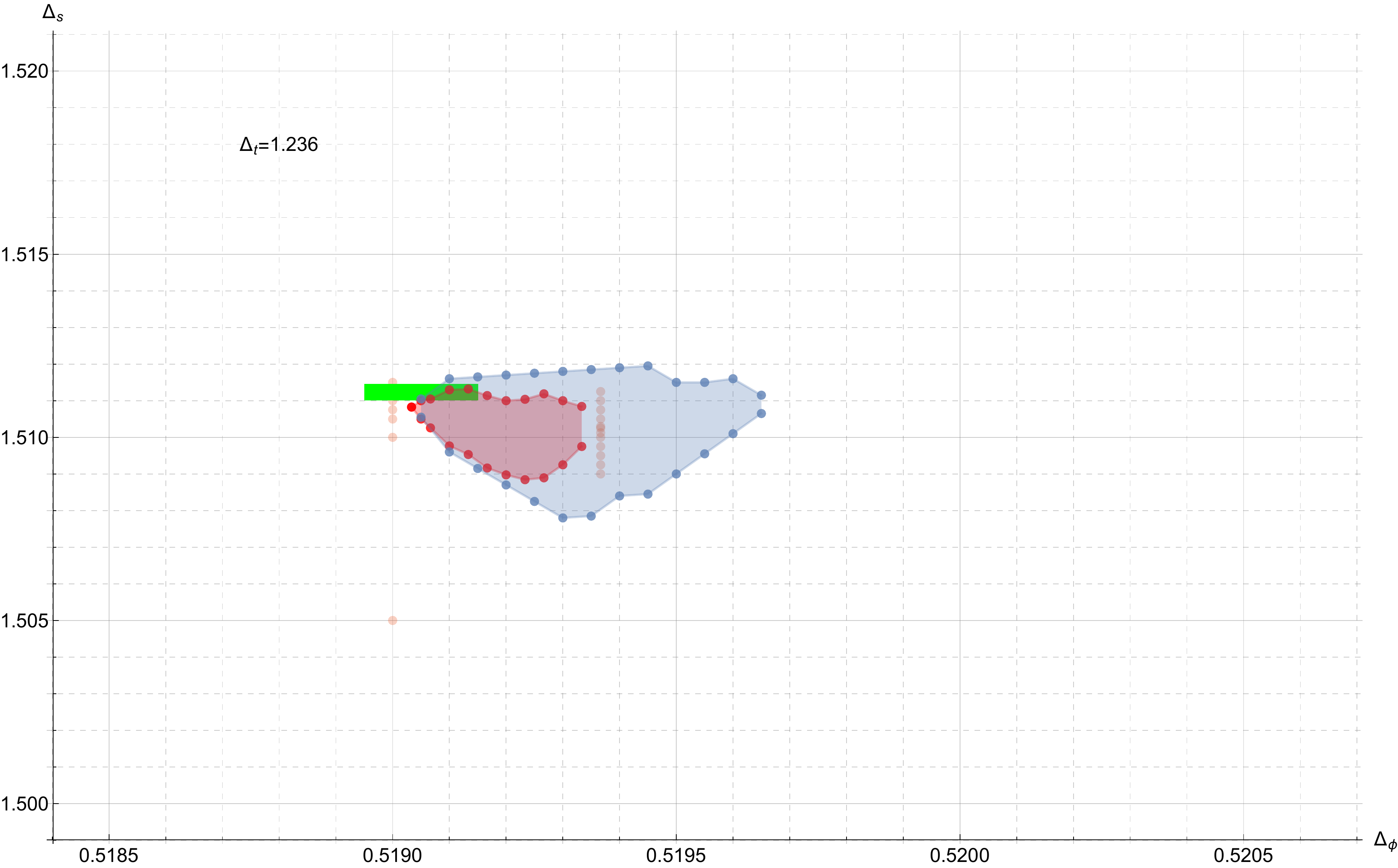}
	\end{minipage}
	\begin{minipage}{0.50\hsize}
		\centering
		\includegraphics[width=\textwidth]{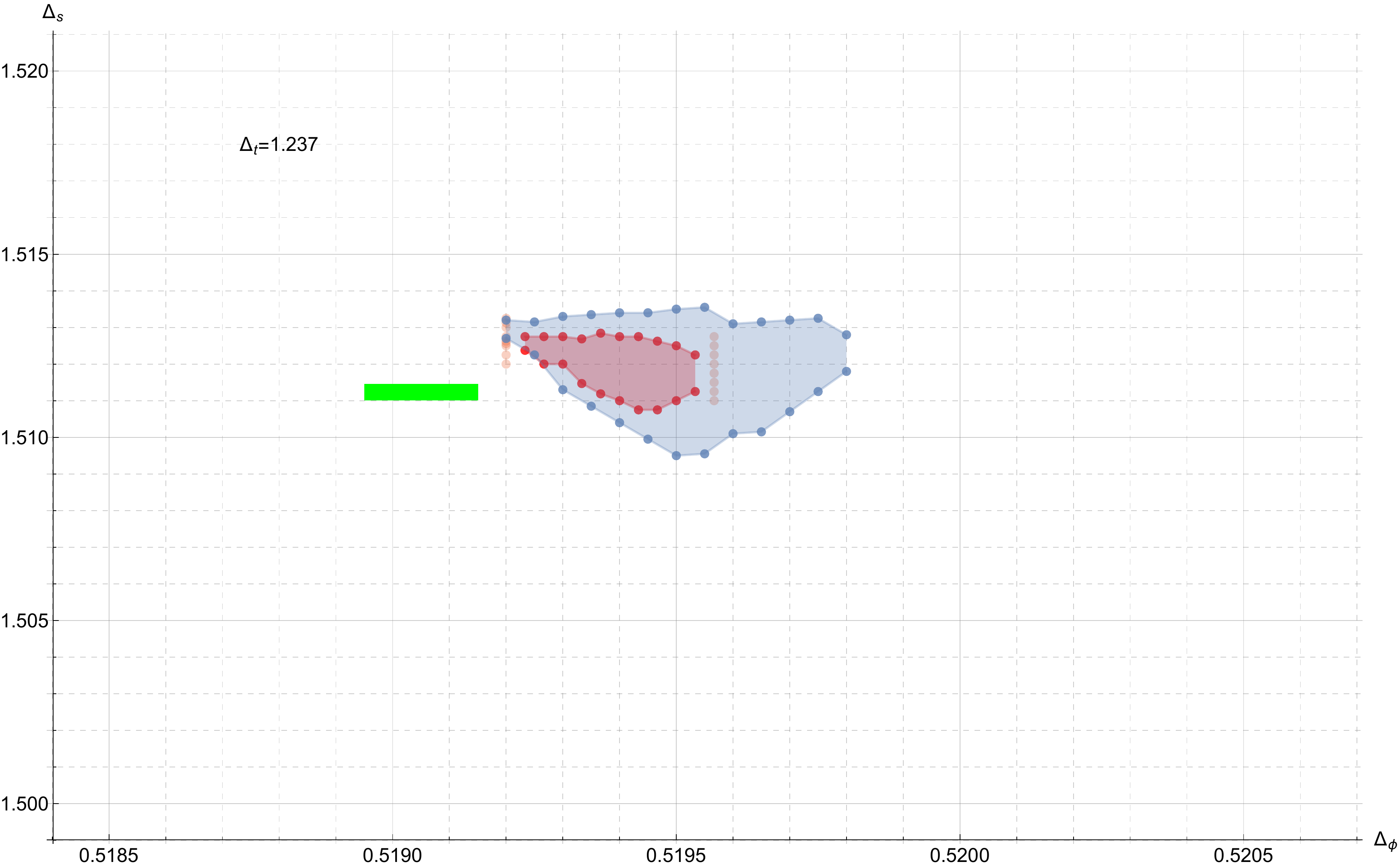}
	\end{minipage}\\
	\begin{minipage}{0.50\hsize}
		\centering
		\includegraphics[width=\textwidth]{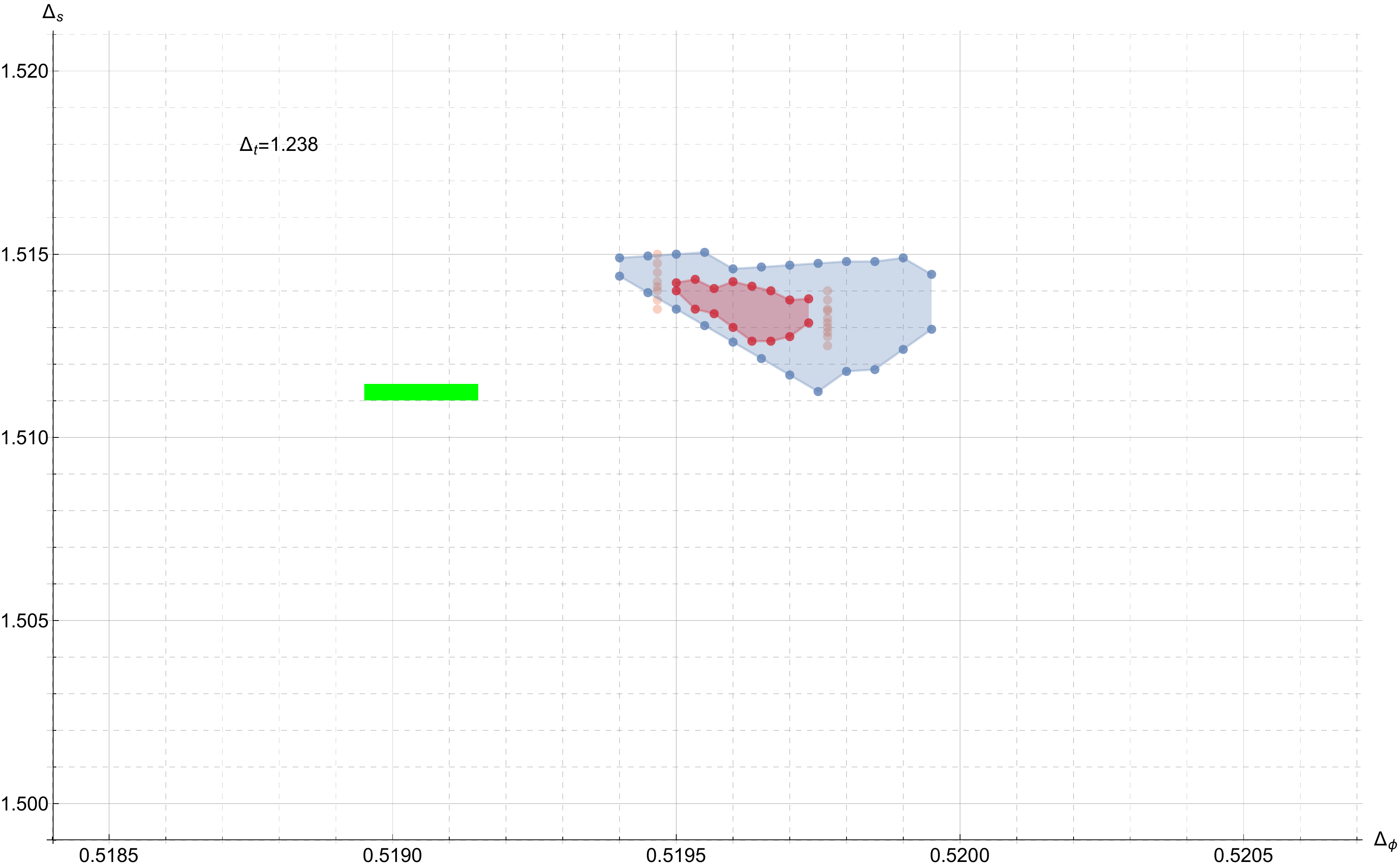}
	\end{minipage}
	\begin{minipage}{0.50\hsize}
		\centering
		\includegraphics[width=\textwidth]{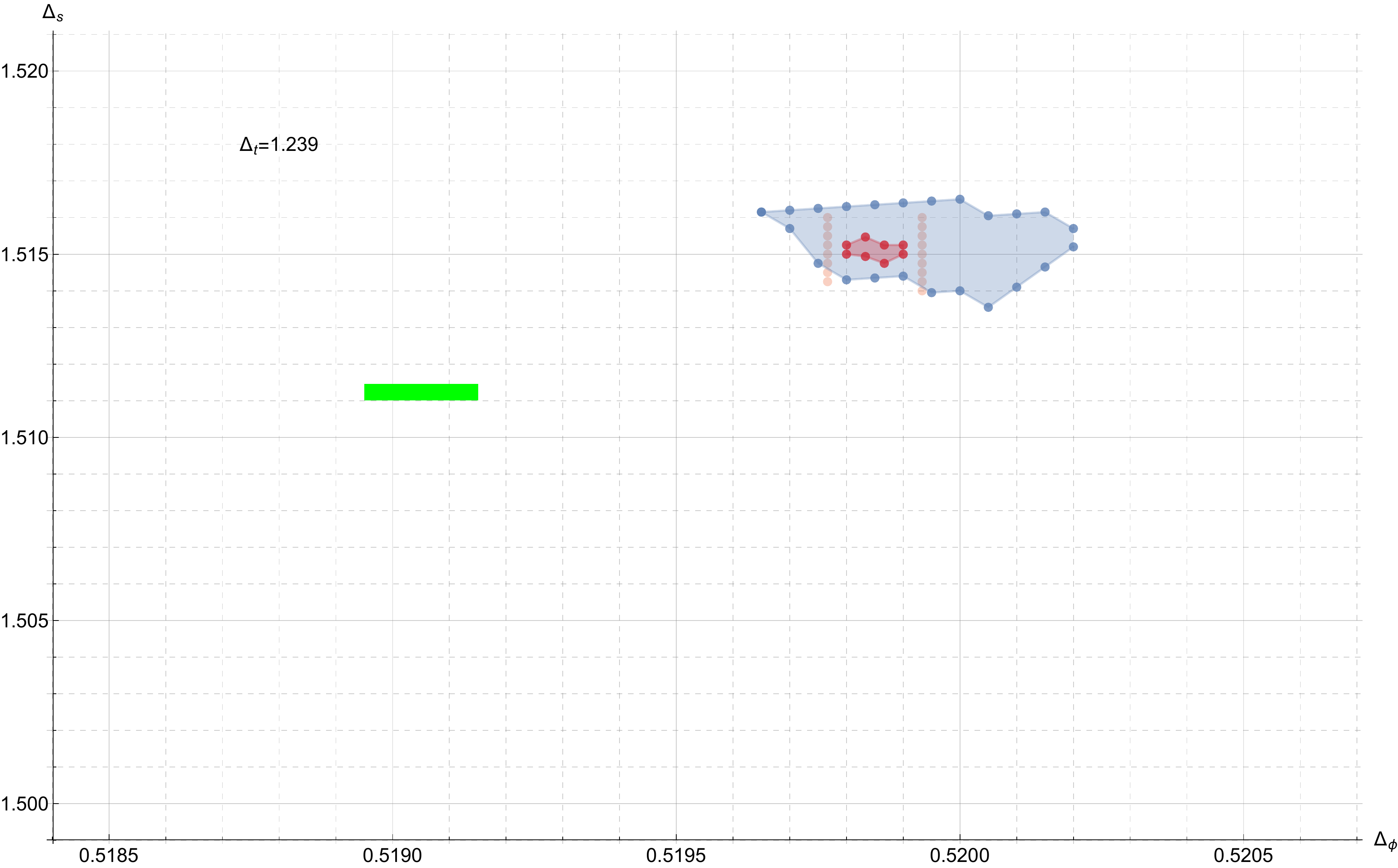}
	\end{minipage}
\end{tabular}
\caption{The island from the mixed correlator bootstrap of the $O(2)$ model.
Here, the red regions are our results at $\Lambda=25$;
the blue regions are those obtained in Appendix B of \cite{Kos:2015mba};
and the green rectangle shows the Monte-Carlo results of \cite{Campostrini:2006ms}.
\label{fig:o2}}
\end{figure}

\begin{figure}
	\centering
	\includegraphics[width=\textwidth]{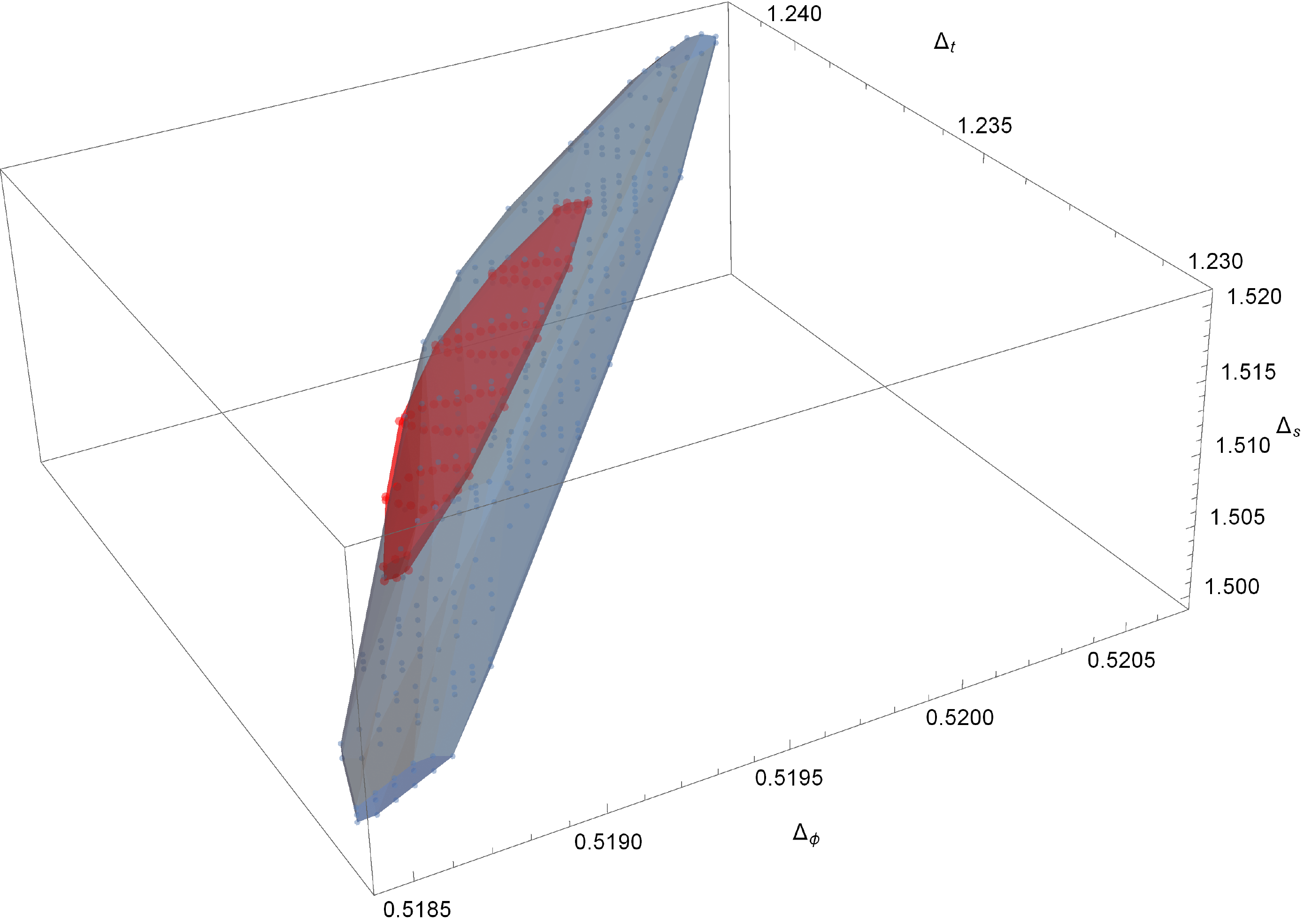}
	\caption{Three-dimensional plot of the $O(2)$ island.
	The blue region is the one obtained in Appendix B of \cite{Kos:2015mba},
	and the red region is our result.
	\label{fig:o2-3d}}
\end{figure}

We used $\Lambda=25$ and
obtained the island in the $(\Delta_s,\Delta_\phi,\Delta_t)$ space shown in Fig.~\ref{fig:o2} and Fig.~\ref{fig:o2-3d},
where the results from Appendix B of \cite{Kos:2015mba} are also presented.\footnote{%
The authors thank the authors of \cite{Kos:2015mba}, in particular David Simmons-Duffin,
for providing the raw data used to create their original figures to be reproduced here.
The authors also thank Shai Chester and Alessandro Vichi for helpful discussions on the computations.}
Our bound is the following:
\begin{equation}
1.50597 \le \Delta_s \le 1.51547,\qquad
0.5188 \le \Delta_\phi \le 0.5199,\qquad
1.234 \le \Delta_t \le 1.239.
\end{equation}

\section*{Acknowledgements}

The authors thank Shai Chester, Walter Landry, David Simmons-Duffin, and Alessandro Vichi for helpful comments on an early version of the manuscript,
and Tomoki Ohtsuki for discussions on \code{cboot}.
The authors also thank Yu Nakayama for spotting crucial typos in the v1 of the manuscript which prevented the sample codes to actually run correctly.
MG is partially supported by the Leading Graduate Course for Frontiers of Mathematical Sciences and Physics.
YT is partially supported by JSPS KAKENHI Grant-in-Aid (Wakate-A), No.17H04837
and JSPS KAKENHI Grant-in-Aid (Kiban-S), No.16H06335,
and also by WPI Initiative, MEXT, Japan at IPMU, the University of Tokyo.

\appendix

\section{Hot-starting the semi-definite programming solver}
\label{sec:hot}

In this appendix, we briefly describe a simple method to often significantly reduce the running time of the semi-definite programming solver during the numerical bootstrap.
For the details of converting the numerical bootstrap into the semi-definite programming, we refer the reader to \cite{Simmons-Duffin:2015qma};
our discussion will be brief.

Recall that in the semi-definite programming we consider maximizing $b_a y^a$
under the condition
\begin{equation}
A_{ij}^u Y^{ij}+B_a^u y^a=c^u,\qquad Y^{ij}\succeq 0
\label{dual}
\end{equation}
where the input data are \begin{equation}
b_a, \quad c^u, \quad B_a^u,\quad A_{ij}^u
\end{equation}
 and we vary \begin{equation}
y^a,\quad Y^{ij}.
\end{equation}
Here, the indices are such that
\begin{equation}
a=1,\ldots,N;\qquad
 u=1,\ldots,P;\qquad
 i,j=1,\ldots,K
\end{equation}
and the indices $i$ and $j$ are assumed to be symmetric.
$Y^{ij}\succeq 0$ means that the matrix $Y$ is positive semi-definite.

This is the \emph{dual} form of the problem, while the \emph{primal} form is that we minimize $c^u x_u$
under the condition
\begin{equation}
X_{ij}= A^u_{ij} x_u,\quad
B_a^u x_u=b_a,\quad
X_{ij}\succeq 0
\label{primal}
\end{equation}
where we have the same input data as above and we vary \begin{equation}
x_u, \quad X_{ij}.
\end{equation}

When $(x,X)$ or $(y,Y)$ satisfies the respective equality condition \eqref{primal} or \eqref{dual}, they are called primal or dual feasible.
The duality gap defined as $c^u x_u-b_a y^a$ is guaranteed to be non-negative for a primal feasible $(x,X)$ and a dual feasible $(y,Y)$.
When the duality gap vanishes, both $(x,X)$ and $(y,Y)$ satisfy the respective optimization problems, and $XY=0$.
A semi-definite program solver starts from an initial point $(x,X,y,Y)$, which is allowed not to satisfy the equality constraints in \eqref{dual} and \eqref{primal}, and update the values of $(x,X,y,Y)$ via a generalized Newton search so that they become feasible up to an allowed numerical error we specify.

In the application to the numerical bootstrap, the bootstrap constraints are turned into a maximization problem of the dual form discussed above.
The aim is to construct an exclusion plot of the scaling dimensions $\Delta_{1,\ldots,n}$ of external operators $\phi_{1,\ldots,n}$.
Depending on the precision we want to impose, we pick a fixed value of $K,N,P$,
and we construct $c^u$, $B^u_a$, $A_{ij}^u$ as a function of $\Delta_{1,\ldots,n}$.
We often simply set $b=0$ and look for a dual feasible solution.
If one is found, the chosen set of values $\Delta_{1,\ldots,n}$ is excluded.
To construct an exclusion plot, we repeat this operation for many sets of values $\Delta_{1,\ldots,n}$.

In the existing literature, and in the sample implementations available in the community,
the semi-definite program solver is often repeatedly run with the initial value $(x,X,y,Y)\propto (0,\Omega_P I_{K\times K},0,\Omega_D I_{K\times K})$ where $I_{K\times K}$ is the unit matrix and $\Omega_{P,D}$ are real constants.
Our improvement is simple and straightforward: for two sets of nearby values $\Delta_{1,\ldots,n}$ and $\Delta'_{1,\ldots,n}$, we reuse the final value $(x_*,X_*,y_*,Y_*)$ for the previous run as the initial value for the next run.
For nearby values of $\Delta_{1,\ldots,n}$, the updates of the values $(x,X,y,Y)$ via the generalized Newton search are expected to follow a similar path.
Therefore, we can expect that reusing the values of $(x,X,y,Y)$ might speed up the running time, possibly significantly.
We call this simple technique the hot-starting of the semi-definite solver.
For this purpose, we implemented a new option \code{--initialCheckpointFile} to \code{sdpb}, so that the initial value of $(x,X,y,Y)$ can be specified at the launch of \code{sdpb}.
The code has been merged to the \code{master} branch of \url{https://github.com/davidsd/sdpb}.

We have not performed any extensive, scientific measurement of the actual speedup by this technique.
But in our experience, the \code{sdpb} finds the dual feasible solutions about 10 to 20 times faster than starting from the default initial value.

There are a couple of points to watch out in using this technique:
\begin{itemize}
\item In the original description of \code{sdpb} in \cite{Simmons-Duffin:2015qma}, it is written in Sec.~3.4 that
\begin{quotation}
In practice, if \code{sdpb} finds a primal feasible solution $(x,X)$ after some number of iterations, then it will never eventually find a dual feasible one. Thus, we additionally include the option \code{--findPrimalFeasible}
\end{quotation}
and that finding a primal feasible solution corresponds to the chosen set of values $\Delta_{1,\ldots,n}$ is considered allowed.
This observation does not hold, however, once the hot-start technique is applied.
We indeed found that often a primal feasible solution is quickly found, and then a dual feasible solution is found later.
Therefore, finding a primal feasible solution should not be taken as a substitute for never finding a dual feasible solution.
Instead, we need to turn on options \code{--findDualFeasible} and \code{--detectPrimalFeasibleJump} and turn off \code{--findPrimalFeasible}.\footnote{%
Walter Landry pointed out that our observation here seems to be related to the bug in \code{sdpb}, where the primal error was not correctly evaluated. 
This bug is corrected in the \code{sdpb} version 2 in the \code{elemental} branch, released in early March 2019.
}
\item
From our experiences, it is useful to \emph{prepare} the tuple $(x,X,y,Y)$
by running the \code{sdpb} for two values of $\Delta_{1,\ldots,n}$,
such that one is known to belong to the rejected region
and another is known to belong to the accepted region, so that the tuple $(x,X,y,Y)$
\emph{experiences} both finding of a dual feasible solution
and detecting of a primal feasible jump.
Somehow this significantly speeds up the running time of the subsequent runs.
\item
When one reuses the tuple $(x,X,y,Y)$ too many times,
the control value $\mu$ which is supposed to decrease sometimes mysteriously starts to increase.
At the same time, one observes that the primal and dual step lengths $\alpha_P$ and $\alpha_D$ (in the notation of \cite{Simmons-Duffin:2015qma}) become very small.
This effectively stops the updating of the tuple $(x,X,y,Y)$.
When this happens, it is better to start afresh,
or to reuse the tuple $(x,X,y,Y)$ from some time ago which did not show this pathological behavior.
\end{itemize}

\bibliographystyle{ytphys}
\baselineskip=.95\baselineskip
\bibliography{ref}

\end{document}